\documentstyle{mn}

\newcounter{parentequation}

\def\ltsima{$\; \buildrel < \over \sim \;$}
\def\gtsima{$\; \buildrel > \over \sim \;$}
\def\simlt{\lower.5ex\hbox{\ltsima}}
\def\simgt{\lower.5ex\hbox{\gtsima}}

\def\etal{{\it et al.}\rm}
\def\etals{{\it et al. }\rm}
\def\mk2{\mu {\rm K}^2}

\begin{document}

\title[Statistics of CMB Quadrupole]
{The Statistical Significance of the  Low CMB Mulitipoles}

\author[G. Efstathiou]{G. Efstathiou\\
Institute of Astronomy, Madingley Road, Cambridge, CB3 OHA.}

\maketitle

\begin{abstract}
The Wilkinson Microwave Anisotropy Probe (WMAP) has measured lower
amplitudes for the temperature quadrupole and octopole anisotropies
than expected in the best fitting (concordance) $\Lambda$-dominated
cold dark matter ($\Lambda$CDM) cosmology. Some authors have argued
that this discrepancy may require new physics. Yet the statistical
significance of this result is not clear. Some authors have applied
frequentist arguments and claim that the discrepancy would occur by
chance about $1$ time in $700$ if the concordance model is
correct. Other authors have used Bayesian arguments to claim that the
data show marginal evidence for new physics. I investigate these
confusing and apparently conflicting claims in this paper using a
frequentist analysis and a simplified Bayesian analysis. On either
analysis, I conclude that the WMAP results are consistent with the
concordance $\Lambda$CDM model.

\vskip 0.1 truein


\vskip 0.3 truein

\end{abstract}

\section{Introduction}

The WMAP satellite  (Bennett \etals 2003a; Hinshaw \etals 2003;
Spergel \etals 2003) has led to a precise measurement of 
the temperature anisotropy power spectrum, $C_\ell$,  from multipoles
$\ell =2$ to $\ell \sim 600$. The observed temperature power spectrum is
in striking agreement with the predictions of the `concordance'
inflationary $\Lambda$CDM cosmology 
 with parameters consistent with those inferred from
observations made prior to WMAP (compare, for example, 
Wang \etals 2002 and  Spergel \etals 2003).

However, as pointed out by the WMAP team there may be a discrepancy
between the predictions of the $\Lambda$CDM models and the
observations at low multipoles. A low amplitude of the CMB quadrupole
was first found by COBE (Hinshaw \etals 1996), but the new WMAP
observations have led to a more accurate measurement and to tighter
control of systematic errors caused by  residual foreground
emission from the Galaxy. The amplitude of the octopole
measured by WMAP is also low compared to the best fitting
$\Lambda$CDM model and the temperature autocorrelation function
$C(\theta)$ shows  an almost complete lack of signal on angular
scales $\simgt 60$ degrees. Spergel \etals (2003, hereafter S03), 
quantify the latter discrepancy by computing the statistic
\begin{equation}
S = \int_{-1}^{1/2} \left [ C(\theta) \right ]^2 \; d\cos\theta
\end{equation}
for a large number of simulated skies generated from the posterior
distribution of the $\Lambda$CDM cosmology. They conclude  that the
probability of finding a value of $S$ smaller than that observed is
about $1.5 \times 10^{-3}$. This low probability, if correct, suggests
a discrepancy between the $\Lambda$CDM cosmology and the observed low
CMB multipoles,  indicating a need for new physics.  Indeed, a number of
authors have explored various models that might reproduce the low
quadrupole and octopole. For example, S03 and Tegmark, de Oliveira
Costa and Hamilton (2003) suggest that the effect might be associated
with the small size of a finite universe, while Efstathiou (2003a) and
Contaldi \etals (2003) have proposed a cut-off in the primordial power
spectrum associated with spatial curvature.  Cline, Crotty and
Lesgourgues (2003) and  Feng and Zhang (2003) consider multi-field
inflation models, while DeDeo, Caldwell and Steinhardt (2003) consider
quintessence models with an equation of state that leads to a partial
cancellation of the usual integrated Sachs-Wolfe effect.  Evidently,
theorists are not short of ideas that might account for the
observations.

But is new physics necessary? Is the probability of $1.5 \times
10^{-3}$ derived in S03 correct, or has the significance of the
discrepancy been overestimated? Do modified models provide
statistically significantly better fits to the data than the
concordance $\Lambda$CDM model?  Some of the recent literature on
these points is confusing. For example,
Bridle \etals (2003), Cline \etals (2003) and Contaldi \etals
(2003)  perform Bayesian analyses of the WMAP data
to test whether the low multipoles require a sharp break in the
primordial spectrum.  Although the data favour a break at a wavenumber
$k_c \sim 3 \times 10^{-4} {\rm Mpc}^{-1}$, the concordance model with
$k_c = 0$ is not strongly excluded.  Is this conclusion compatible
with the SO3 analysis of the $S$ statistic? The questions raised in
this paragraph are addressed in this paper. Tegmark \etals (2003) comment
on a possible alignment between the quadrupole and octopole. This effect
is ignored in this paper, which  focuses exclusively on the
statistical significance of the amplitudes of the quadrupole and octopole.
For an analysis of the statistical significance of alignments,
see de Oliveira-Costa \etals (2003).

\section{Observations and Fiducial Concordance Model}

It has become common to use Monte-Carlo Markov Chains (MCMC) to
evaluate the posterior distributions of cosmological parameters given
observations of the CMB power spectra and their covariance matrices
(see Christensen \etals 2001; Lewis and Bridle 2002; Verde \etals
2003). As a fiducial model, we follow the MCMC
analysis of Bridle \etals (2003) and adopt a
spatially flat $\Lambda$CDM cosmology specified by $6$ parameters, a
constant scalar spectral index $n_s$, spectral amplitude $A_s$, Hubble
constant $h = H_0/(100 {\rm km} {\rm s}^{-1} {\rm Mpc}^{-1})$, baryon
density $\omega_b \equiv \Omega_b h^2$, CDM density $\omega_c \equiv
\Omega_c h^2$ and redshift of reionization $z_{\rm eff}$. Tensor modes
are ignored in this analysis. The input CMB data consists of the WMAP
temperature and temperature-polarization cross-correlation power
spectra (and associated programmes to compute the likelihood function,
see Verde \etals 2003) supplemented with  measurements at higher multipoles
($800 \simlt \ell \simlt 2000$) from CBI (Pearson \etals 2003), ACBAR
(Kuo \etals 2003) and VSA (Grainge \etals 2003).

Figure 1 shows a histogram of the quadrupole amplitudes from a set of
MCMC chains\footnote{The chains have been made available by Antony
Lewis at the following web site http://cosmologist.info/cosmomc/.} for
this six parameter model. The peak occurs at a quadrupole
amplitude of $\Delta T^2_2 \approx 1250\; \mu {\rm K}^2$
($\Delta T^2_\ell \equiv \ell (\ell + 1) C_\ell/(2 \pi)$) and the distribution
is quite narrow;  few samplings have quadrupole amplitudes smaller than
$1000 \; \mu {\rm K}^2$ or greater than $2000 \; \mu {\rm K}^2$.
As a fiducial model, we set $h=0.72$, $n_s = 1.0$, $\omega_b = 0.024$
and $\omega_c = 0.12$ and choose $z_{\it eff}$ so that the optical
depth for Thomson scattering is $\tau = 0.17$. These numbers are very 
close to those that give the maximum likelihood to the data used to generate
Figure 1, but some have been adjusted slightly so that they are
consistent with other data, {\it e.g.} the HST key project measurement
of the Hubble constand (Freedman \etals 2001). The quadrupole and octopole
amplitudes for this fiducial model are $\Delta T^2_2 = 1140 \; \mu {\rm K}^2$
and $\Delta T^2_3 = 1060 \; \mu {\rm K}^2$. To illustrate the sensitivity 
to the parameters of the fiducial model, we will show how various results
change if the quadrupole and octopole amplitude are lowered to 
$\Delta T^2_2 = 1000 \; \mu {\rm K}^2$ and $\Delta T^2_3 = 930 \; \mu {\rm K}^2$,
{\it i.e.} towards the lower end of the allowed range according to Figure 1.

The WMAP quadrupole and octopole amplitudes in the publicly available
data release are given as $\Delta T^2_2 = 123 \; \mu {\rm K}^2$ and
$\Delta T^2_3 = 611\; \mu {\rm K}^2$.  The quadrupole amplitude in
particular (shown by the dashed line in Figure 1) is much lower than
the amplitude of the fiducial $\Lambda$CDM model. The Bennett \etals
(2003a) WMAP summary paper lists the quadrupole amplitude as $154 \pm
70 \; \mu {\rm K}^2$, slightly higher than the value in the public
data release. The error on this number is a 95\% confidence limit on
the uncertainty associated mainly from modelling foreground Galactic
emission. (For comparison, the quadrupole amplitude measured by COBE
is $\Delta T^2_2 = 240\pm^{340}_{124}\; \mu {\rm K}^2$, Hinshaw \etals
1996). Full details of how the WMAP team arrive at this error estimate
have not yet been published, but it would seem to be reasonable since
Tegmark \etals (2003) find a quadrupole amplitude of $202\; \mu {\rm
K}^2$ (including the small $\sim 4 \mk2$ contribution from the
kinematic quadrupole) from an analysis of their all-sky foreground
subtracted WMAP map. Tegmark \etals (2003) find an octopole amplitude
of $866 \; \mu {\rm K}^2$ from their all-sky map.  We adopt these
numbers (which for the quadrupole is at the upper end of the Bennett
\etals error range) to illustrate the effects of systematic
uncertainties associated with contamination by Galactic
emission. Tegmark \etals find similar numbers for an all-sky analysis
of the WMAP internal linear combination foreground subtracted CMB map
(Bennett \etals 2003b).

\begin{figure}

\vskip 2.8 truein

\includegraphics{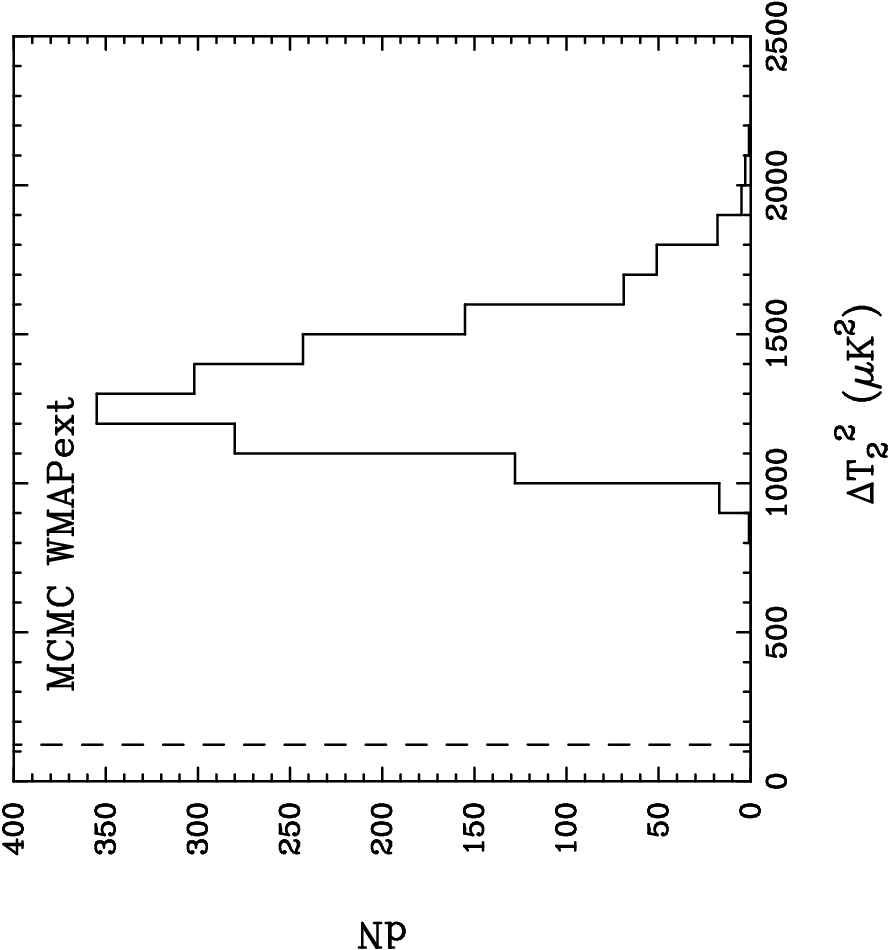}

\caption
{The histogram shows the quadrupole amplitudes for a sample of MCMC
chains for a six parameter $\Lambda$CDM fit to WMAP+CBI+ACBAR+VSA
data. The vertical dashed line shows the quadrupole amplitude observed by
WMAP.}

\label{figure1}

\end{figure}

\section{Frequentist Statistics}


\begin{table*}
\bigskip

\centerline{\bf \ \ \  Table 1a:  
Frequentist estimates of the Quadrupole and Octopole Discrepancy}

\begin{center}

\begin{tabular}{cccccccc} \hline \hline
\smallskip 
$(\Delta T^2_2)^T$ & $(\Delta T^2_3)^T$ & $P(\Delta T^2_2 < 123)$& $P(\Delta T^2_3 < 611)$&
 $\;\;P(\Delta T^2_2 < 123, $ & $P(\Delta T^2_2 < 202)$& 
$P(\Delta T^2_3 < 870)$
& $\;\;P(\Delta T^2_2 < 202,$ \cr 
 & & & & \qquad $\Delta T^2_3 < 611)$ & &  &\qquad  $\Delta T^2_3 < 870)$   \cr
 &  & & & & & &   \cr
$1140$ & $1060$ & $0.013$ & $0.24$  & $0.0032$ & $0.036$ & $0.44$ & $0.016$  \cr
$1000$ & $930$ & $0.017$ & $0.31$  & $0.0054$ & $0.046$ & $0.53$  & $0.025$ \cr
\hline
\end{tabular}
\end{center}

\medskip

\centerline{\bf \ \ \  Table 1b:  
Bayesian estimates of the Quadrupole and Octopole Discrepancy}

\begin{center} 

\begin{tabular}{cccccc} \hline \hline
\smallskip 
$\Delta T^2_2$ & $\Delta T^2_3$ & $P((\Delta T^2_2)^T > 1140)$& $P((\Delta T^2_3)^T > 1060)$
&  $P((\Delta T^2_2)^T > 1000)$& 
$P((\Delta T^2_3)^T > 930)$ \cr 
 & & & & &    \cr
$123$ & $611$ & $0.087$ & $0.45$  
 & $0.10$ & $0.53$   \cr
$202$ & $870$ & $0.16$ & $0.66$  &  $0.19$ & $0.73$   \cr
\hline
\end{tabular}

\end{center}

\begin{quote}

\noindent
Note: Table 1a gives the frequencies that the observed amplitudes
$\Delta T^2_2$ and $\Delta T^2_3$ will be less than the specified
values (expressed in $\mk2$) if the true amplitudes are $(\Delta
T^2_2)^T$ and $(\Delta T^2_3)^T$ (see Section 3).  Table 1b gives the
Bayesian frequencies that the observed values of quadrupole and
octopole amplitudes are drawn for a model with true quadrupole and
octopole amplitudes greater than $(\Delta T^2_2)^T$ and $(\Delta
T^2_3)^T$ (see Section 4).
\end{quote}
\end{table*}

In the absence of a sky cut and instrumental noise, the distribution
of $C_\ell$ estimates in a theory with Gaussian amplitudes $a_{\ell
m}$ follows a $\chi^2$ distribution,
\begin{eqnarray}
dP(C_\ell) &  \propto&   
\left ( { C_\ell \over C^T_\ell} \right)
^{{2 \ell - 1 \over  2}} \rm {exp} \left ( - {(2 \ell + 1) C_\ell
\over 2 C^T_\ell } \right )  \; {dC_\ell \over C^T_\ell},  \label{equ2} 
\end{eqnarray}
where $C_\ell^T$ is the expectation value of $C_\ell$.  Integrating equation 
(\ref{equ2}), the probability of observing 
a value $\le C_\ell$ is given by 
\begin{equation}
P(\le C_\ell) = {\gamma \left ( {2 \ell + 1 \over 2}, {2 \ell+1 \over 2} {C_\ell \over
C^T_\ell} \right )  \over \Gamma \left ( {2 \ell + 1 \over 2} \right )}, \label{equ3}
\end{equation}
where $\gamma$ is the incomplete Gamma function.

In practice, the actual distribution depends on the estimator of
$C_\ell$, the shape of any Galactic cut and, of course, instrumental
noise and other sources of error. Figure 2 shows a histogram of
quadrupole amplitudes determined by applying a pseudo-$C_\ell$
estimator (see {\it e.g.} Hivon \etal, 2002) to a large number of
simulated noise-free maps generated using the power spectrum of the
fiducial $\Lambda$CDM model discussed in the previous section.  The
Kp2 Galactic cut imposed by Hinshaw \etals (2003) was used in the
simulations. Figure 2 shows the resulting distribution of quadrupole
amplitudes, together with a $\chi^2$ distribution (see also
Wandelt,  Hivon, G\'orski,  2001).  The effects of
Galactic cuts on pseudo-$C_\ell$ estimators is discussed in detail by
Efstathiou (2003b, c), however, for the modest Galactic cuts used in the
analysis of WMAP, the quadrupole amplitude is weakly correlated with
higher multipoles and its distribution follows a $\chi^2$ distribution
quite accurately with a variance that is only marginally greater than
the cosmic variance.

The corresponding frequentist statistics are given in Table 1(a) and
agree with equation (2) to within about $50\%$. From this Table, we
see that the probability of observing a quadrupole lower than $123
\mk2$ is about $1.3 \%$ and that the joint probability of finding
quadrupole and octopole amplitudes smaller than those observed is
about $0.32\%$ which is about twice the value inferred by SO3 from
their analysis of the $S$ statistic. In fact, SO3 compare the $S$
statistic with simulations generated from their MCMC chains. If we
rescale the quadrupole and octopole amplitudes of our simulations so
that the amplitudes follow the MCMC distribution plotted in Figure 1,
the joint probability for the quadrupole and octopole amplitudes drops
from $0.32\%$ to $0.21\%$ only slightly larger than the value of
$0.15\%$ deduced by SO3 from the $S$ statistic. Given that the
integration range of the $S$ statistic was chosen {\it a posteriori},
it is not suprising that SO3 find a slightly more significant
discrepancy. The main conclusion to draw from this analysis is that
the significance level deduced by SO3 from the $S$ statistic is
understandable; a similar significance level is deduced from the
quadrupole and octopole amplitudes alone. Most of the weight in the
$S$ statistic is coming from the quadrupole and octopole amplitudes
and any `{\it a posteriori} bias' in the statistic is small.\footnote{
We note here that Gazta\~naga \etal, 2003, find much higher
probabilities by applying a frequentist statistic to estimates of
$C(\theta)$ from WMAP using a conservative Galactic cut. Their results
are puzzling, however, because: (i) they find that their statistic is
extremely sensitive to the size and shape of the Galactic cut; (ii)
their results are inconsistent with the pseudo-$C_\ell$ estimator
used by the WMAP team and the quadratic maximum likelihood
estimator used by Efstathiou (2003c).}

\begin{figure}

\vskip 2.8 truein

\includegraphics{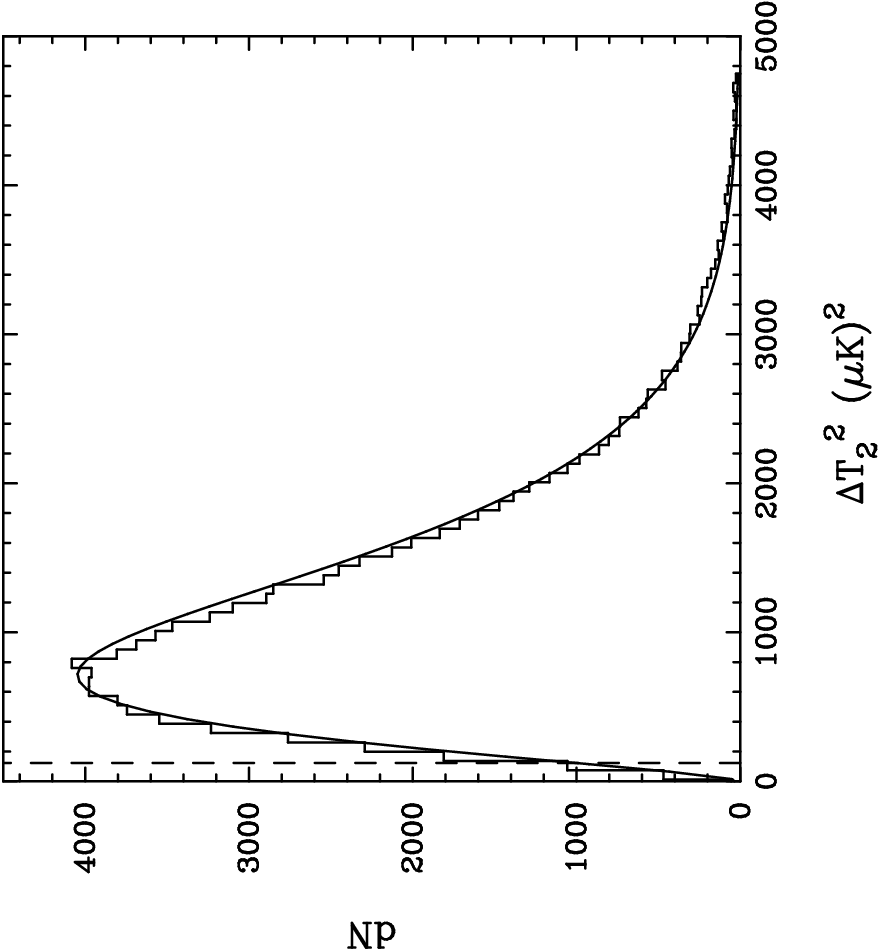}

\caption
{Histogram of quadrupole amplitudes estimated using a pseudo-$C_\ell$
estimator applied to $10^5$ simulations with a Galactic cut of $\pm 10^\circ$
centred on the Galactic plane. The power spectrum of the fiducial $\Lambda$CDM
model discussed in Section 2 was used to generate the simulations. The solid
line shows a $\chi^2$ distribution (equation 2) and the dashed line shows the
quadrupole amplitude measured by WMAP.}

\label{figure2}

\end{figure}

\begin{figure*}

\vskip 3.0 truein

\includegraphics{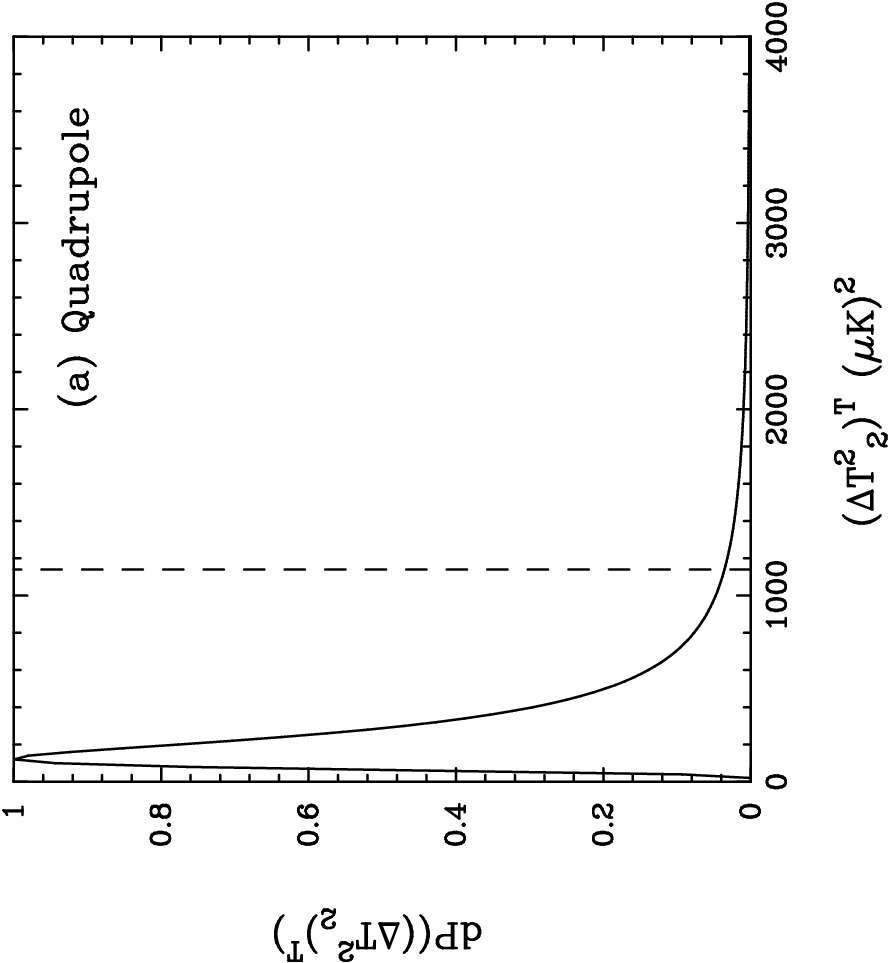}

\includegraphics{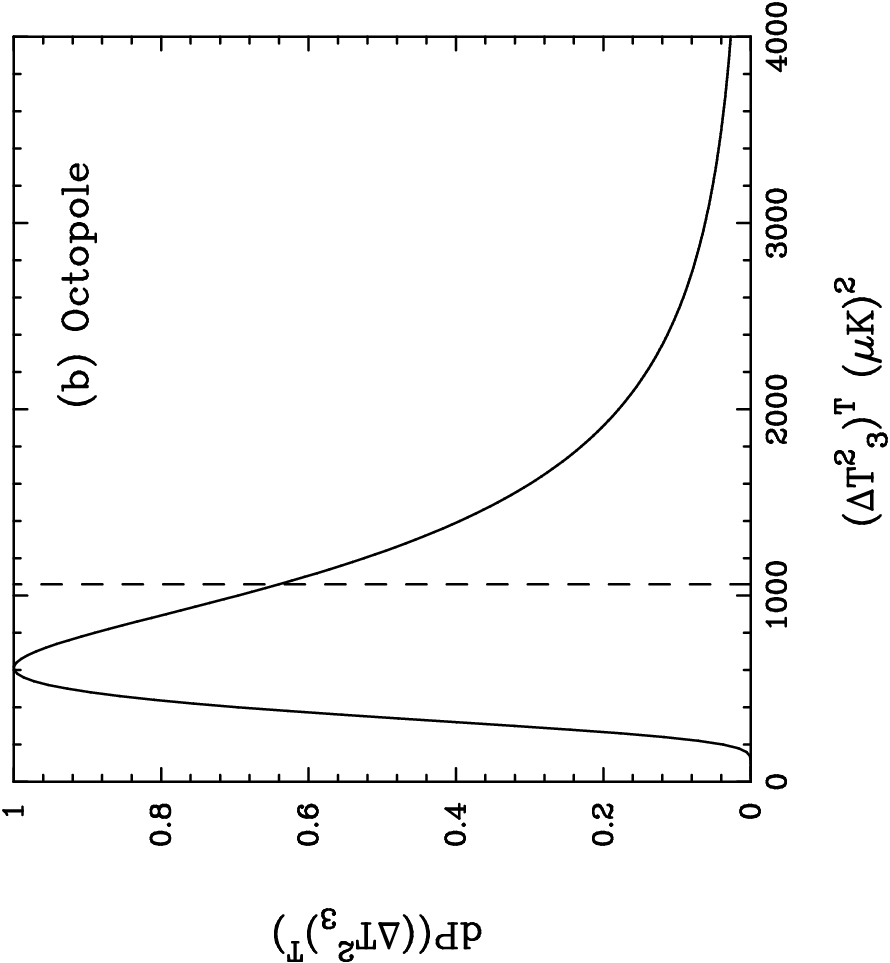}

\caption
{Unnormalised Bayesian probability distributions for the quadrupole and octopole
amplitudes given by equation (\ref{B2}). The dashed lines show the values for
the fiducial $\Lambda$CDM model discussed in Section 2.  }

\label{figure3}

\end{figure*}

 Table 1a also lists the frequentist probabilities if the
quadrupole and octopole amplitudes are $202 \mk2$ and $870 \mk2$
respectively, which we have argued in Section 2 are within the allowed
range given uncertainties in subtracting Galactic
foregrounds. (The analysis of the WMAP temperature-polarization
power spectrum, $C^{TE}$, shows clear evidence for a Galactic
polarized signal (Kogut \etals 2003). Since the systematic errors in $C^{TE}$ from
Galactic emission have not yet been quantified fully, and since the
randon errors are large, we ignore the WMAP $C^{TE}$ measurements in
the rest of this paper.) The joint probability of finding quadrupole
and octopole amplitudes below these values given the fiducial
$\Lambda$CDM model is $1.6\%$, {\it i.e.} about $5$ times greater than
the value deduced for quadrupole and octopole amplitudes of $123 \mk2$
and $611 \mk2$, an order of magnitude greater than deduced by SO3 from
the $S$ statistic. If the parameters of the $\Lambda$CDM model are
adjusted to give values for the quadrupole and octopole amplitudes that
lie towards the lower end of the allowed range, the joint probability
for the quadrupole and octopole rises to $2.5\%$. We conclude that it
is premature to rule out the $\Lambda$CDM cosmology because of the low
quadrupole and octopole amplitudes.  The true values of the quadrupole
and octopole for our CMB sky have not yet been determined with a
sufficiently high accuracy, and the theoretical expectations of the
$\Lambda$ cosmology are not constrained tightly enough, to exclude the
model at a high significance level. The S03 analysis of the $S$
statistic exaggerates the statistical significance of any discrepancy
with the $\Lambda$CDM model because they did not consider
errors in the autocorrelation function arising from
innacurate subtraction of  Galactic emission.

\section{Bayesian Statistics}

As mentioned in the Introduction, a number of papers ({\it e.g.}
Bridle \etals 2003, Contaldi \etals 2003, Cline \etals 2003) have
applied Bayesian methods to analyse more complex models, for example,
$\Lambda$CDM models with a sharp break or truncation in the initial power spectrum on
large scales. These papers ignore the possible systematic errors in
the WMAP power spectra discussed above, yet even so they report no
strong evidence for the introduction of any additional parameters. How
is this conclusion compatible with the SO3 analysis of the $S$
statistic or the simple frequentist tests described in the
previous Section? In this Section we provide an answer by applying
Bayes' theorem to an intentionally simple model.

According to Bayes' theorem, the posterior probability 
of hypothesis $H$   given the data $D$ is 
\begin{equation}
   P(H \vert D) \propto P(D \vert H) P(H), \label{B1}
\end{equation}
where $P(D \vert H)$ is the probability of the data $D$ given $H$
and $P(H)$ is the prior probability  of $H$. Let us adopt the 
hypothesis that the true amplitude ${C_\ell^T}$
lies in the range $C_\ell^T \pm dC_\ell^T$, then applying
 equation (\ref{B1}) 
and assuming a uniform prior for $C_\ell^T$, the posterior 
probability distribution for $C_\ell^T$ is
\begin{equation}
   dP(C_\ell^T) \propto  { 1 \over (C_\ell^T)^{2 \ell + 1 \over 2}}
\rm {exp} \left ( - {(2 \ell + 1) C_\ell
\over 2 C^T_\ell } \right )  \; {dC^T_\ell},  \label{B2} 
\end{equation}
where $C_\ell$ is the observed amplitude. Equation (\ref{B2}) is 
proportional to the likelihood function, which has its maximum value
at $C^T_\ell = C_\ell$. The distributions (\ref{B2}) are plotted for
the quadrupole and octopole in Figure 3.

The distributions plotted in Figure 3 give the posterior probabilities
that the true amplitudes $C_\ell^T$ take on any particular value and
so we can use these figures to test how `disconnected' the observed
values of $C_2$ and $C_3$ (corresponding to the
peaks of the probability distributions) are from those of the fiducial model
(indicated by the vertical dashed lines). The ratio of these
probabilities is $p(C_2)/p((C_2^T)_{\rm fid} = 28$ and
$p(C_3)/p((C_3^T)_{\rm fid} = 1.6$; neither of these ratios is high
and so we conclude that the observed amplitudes do not provide
strong evidence against the fiducial $\Lambda$CDM model.  

Why do these numbers indicate a weaker rejection of the model than the
frequentist statistics of Table 1a?  Let us recast the Bayesian
analysis in frequentist language. Imagine that we draw values of
$(C^T_\ell)$ from a uniform distribution between $0$ and an upper
limit $(C^T_\ell)_{\rm max}$ (the exact value of this maximum limit is
unimportant as long as it extends well into the tail of the
distribution (\ref{equ2})). For each draw, generate a random value of
$C_\ell$ from the $\chi^2$ distribution (\ref{equ2}) and for those
values that lie within a narrow interval around the observed value of
$C_\ell$ compute the frequency with which $C_\ell^T$ exceeds a critical
value $(C_\ell^T)_{\rm crit}$.  This frequency is just the integral
over the probability distribution (\ref{B2})
\begin{equation}
   P\left ( C_\ell^T > (C_\ell^T)_{\rm crit} \right)  = \int_{(C_\ell^T)_{\rm crit}}^\infty dP(C_\ell^T),
  \label{B3} 
\end{equation}
(where the upper limit $(C^T_\ell)_{\rm max}$ has been replaced by
infinity). Numerical values for these frequencies for the octopole and
quadrupole (neglecting minor effects from a cut sky) are given in
Table 1b for values of $(C_\ell^T)_{\rm crit})$ equal to those of
the fiducial model and for values at the low end of the range found
from the MCMC chains. The latter numbers are the more useful because
if these frequencies turn out to be low, then there is little overlap
between the posterior distributions of Figure 3 and the distributions
of quadrupole and octopole amplitudes from the MCMC chains. This would
force us to reject the concordance $\Lambda$CDM model.

However, we find that the frequency with which $(\Delta^T_2)^2 >
1000\mk2$, given the observed WMAP quadrupole of $123 \mk2$ is only
$0.10$, and so again we conclude that the evidence against the
$\Lambda$CDM model is marginal. Of course, this test is different to
the frequentist test discussed in Section 3 (Table 1a), but it is easy
to understand why the two tests give different impressions of a discrepancy.
We can see from Figure 2 that the probability
of finding a quadrupole amplitude as low as that observed, given the
fiducial model, is not improbably small and so if we assume a uniform
prior for $C^T_2$, low values of $C^T_2$ simply do not have enough weight
to exclude the quadrupole amplitude of the fiducial model
at high significance.

Since $C^T_2$ varies from zero to infinity, should we not have used
Jeffreys' prior (Jeffreys 1939, Jaynes 2003), $dC^T_2/C^T_2$, thus
giving extra weight to low values of $C^T_2$? No, because there is a
natural scale for $C^T_2$, namely the amplitude of the fiducial model
$(C^T_2)_{\rm fid}$. In any reasonable physical model, it is
impossible to get a quadrupole amplitude that is very much smaller
than that of the fiducial model because $C_2$ involves an integral of
the perturbation spectrum over wavenumber $k$. Since there is strong
evidence in favour of the fiducial model for wavenumbers $k \simgt
10^{-3} {\rm Mpc}^{-1}$ (Bridle \etals 2003) we should strongly
disfavour models with very low values of $C^T_2$.  The assumption of a
uniform prior over the range, say, $\sim 10^{-1} (C^T_2)_{\rm fid}$ to
a few times $(C^T_2)_{\rm fid}$ is physically reasonable and
relatively benign, although (as with all of the Bayesian analyses
referred to in this paper) we must recognise that there is some
dependence of the posterior probabilities on the form of the prior.

In conclusion, the Bayesian frequencies given in Table 1b provide a
meaningful comparison of the fiducial $\Lambda$CDM model to the WMAP
data and they indicate marginal evidence for any discrepancy.  In the
opinion of this author, the Bayesian analysis is preferable to the
frequentist analysis of Section 3 which is, in any case,  inconclusive
because of systematic errors in the quadrupole and octopole
amplitudes. The Bayesian frequencies listed in Table 1b could be
misleading only if there is persuasive evidence that the priors on
$C^T_2$ and $C^T_3$ should be strongly skewed towards much smaller
values than those of the fidicual model, in which case the low
amplitudes observed by WMAP add little new information.

\section{Conclusions}

\smallskip

\noindent
{\it (i) Do the WMAP measurements of the quadrupole and octopole
amplitudes conflict with the $\Lambda$CDM cosmology?} Based on the quadrupole
and octopole amplitudes, the answer is
unambiguously no.  The frequentist tests discussed in Section 3 are
inconclusive because there are significant systematic errors in the
WMAP quadrupole and octopole amplitudes.  These errors were neglected
in SO3's analysis of the $S$ statistic and hence their estimate of a
$1$ in $700$ chance of reproducing the observations according to the
concordance $\Lambda$CDM model is an overestimate of the true
odds. The Bayesian analysis of Section 4 suggests that a more
reasonable estimate of the odds is more like $1$ in $10$ or $1$ in
$20$.   Whatever your statistical orientation, there is no 
convincing evidence for a discrepancy with the
concordance   $\Lambda$CDM model.

\smallskip

\noindent
{\it (ii) Do the WMAP measurements of the quadrupole and octopole
amplitudes require new physics?} Despite point (i) above, the
likelihood functions plotted in Figure 3 peak at lower values than
those of the fiducial $\Lambda$CDM model and so will favour models
which predict low quadrupole and octopole amplitudes, provided that
the number of extra parameters required to describe the models is not
too large\footnote{ This condition can be quantified using Bayesian
methods, {\it e.g.} by computing Occam factors (Jaynes 2003, Chapter
20) or Bayesian evidence (see {\it e.g.} Saini, Weller and Bridle
2003).} As an example, consider the analysis of Bridle \etals (2003)
of a $\Lambda$CDM model with an initial spectrum truncated sharply
below a wavenumber $k=k_c$ (see their Figure 2). The WMAP data favour
a truncation at $k_c \sim 3 \times 10^{-4} {\rm Mpc}^{-1}$, thus
favouring new physics, but (consistent with the results of this paper)
a model with $k_c =0$ is not strongly excluded. We conclude that the
WMAP data certainly warrant exploration of models incorporating new
physics, but these models had better make other testable predictions
if they are ever to be strongly preferred  over the concordance
$\Lambda$CDM model.

\smallskip

\noindent
{\it (iii) Can measurements of the low CMB multipoles be improved?} As mentioned
in Section 2, Bennett \etals (2003a) quote an error on the quadrupole amplitude
of $\pm 70 \mk2$ and state that this is  caused largely by errors in subtracting
foreground emission. This error estimate is consistent with the difference in the
quadrupole amplitude measured by Tegmark \etals (2003), who use a different method
to subtract Galactic emission. It may, therefore, be possible to improve on the
accuracy of the quadrupole, and other low CMB multipoles, by applying better methods
of foreground subtraction. 

More accurate estimates of the low multipoles can be obtained by
applying an optimal estimator (see {\it e.g.} Tegmark 1997) rather than the
pseudo-$C_\ell$ estimator used by the WMAP team.  In the
noise-free limit (a good approximation for WMAP on large angular
scales), an optimal estimator will return almost the {\it exact}
values of low multipoles on the cut sky, provided that the sky-cut is
not too large. An
analysis of this sort might establish whether $\Delta T_2^2$ is closer
to $100 \mk2$ or $200 \mk2$, which would be useful, though as
explained in Section 4, the Bayesian analysis is not particularly
sensitive to variations of this magnitude.\footnote{Such an analysis
has been completed since this paper was submitted for publication
(Efstathiou 2003c). The results suggest a quadrupole amplitude of
about $200 \mk2$.}

\medskip

\noindent
{\bf Acknowledgments:} I thank members of the Leverhulme Cosmology
Group in Cambridge, and particularly Sarah Bridle, for useful
discussions. I thank Sarah Bridle and Antony Lewis for supplying the
MCMC chains used in Figure 1.

\end{document}